\documentclass[aps,twocolumn,showpacs,superscriptaddress,groupedaddress,nofootinbib]{revtex4}

\usepackage{graphicx}  
\usepackage{dcolumn}   
\usepackage{bm}        
\usepackage{amssymb}   
\usepackage{amsmath}   
\usepackage{multirow}
\usepackage{hyperref}
\usepackage{braket}
\usepackage{subfigure}
\usepackage{array}
\usepackage[math]{cellspace}
\usepackage[T1]{fontenc}

\usepackage{color}
\usepackage[normalem]{ulem}

\newcommand{\NC}{N_c}

\begin{document}


\title{Dissecting the $\Delta I=1/2$ rule at large $N_c$}

\date{\today}

\author{Andrea Donini}
\affiliation{IFIC (CSIC-UVEG), Edificio Institutos Investigaci\'on, 
Apt.\ 22085, E-46071 Valencia, Spain}
\author{Pilar Hern\'andez}
\affiliation{IFIC (CSIC-UVEG), Edificio Institutos Investigaci\'on, 
Apt.\ 22085, E-46071 Valencia, Spain}
\author{Carlos Pena}
\affiliation{Departamento de F\'{\i}sica Te\'orica and Instituto de F\'{\i}sica Te\'orica UAM-CSIC,
Universidad Aut\'onoma de Madrid, E-28049 Madrid, Spain}
\author{Fernando Romero-L\'opez}
\affiliation{IFIC (CSIC-UVEG), Edificio Institutos Investigaci\'on, 
Apt.\ 22085, E-46071 Valencia, Spain}

\preprint{IFIC/16-39}
\preprint{IFT-UAM/CSIC-16-063}
\preprint{FTUAM-16-26}

\begin{abstract}
We study the scaling of kaon decay amplitudes with the number of colours, $N_c$, in a theory with four degenerate flavours, $N_f=4$. In this scenario,
two current-current operators, $Q^\pm$, mediate $\Delta S=1$ transitions, such as the two isospin amplitudes of non-leptonic kaon decays for $K\to (\pi\pi)_{I=0,2}$, 
$A_0$ and $A_2$. In particular, we concentrate on the simpler $K\to\pi$ amplitudes, $A^\pm$, mediated by these two operators. A diagrammatic analysis of the large-$N_c$ scaling of these observables is presented, which demonstrates the anticorrelation of the leading ${\mathcal O}(1/N_c)$ and ${\mathcal O}(N_f/N_c^2)$ corrections in both amplitudes. Using our new $N_f=4$ and previous quenched data, we confirm this expectation and show that these corrections are {\it naturally} large and may be at the origin of the $\Delta I=1/2$ rule.  The evidence for the latter is indirect, based on the matching of the amplitudes
to their prediction in Chiral Perturbation Theory, from which the LO low-energy couplings of the chiral weak Hamiltonian, $g^\pm$, can be determined. 
A NLO estimate of the $K \to (\pi\pi)_{I=0,2}$ isospin amplitudes
can then be derived, which is in good agreement with the experimental value. 
\end{abstract}
\maketitle

\section{Introduction}

Significant progress has been achieved recently in the lattice determination of $K \rightarrow (\pi\pi)_{I=0,2}$ amplitudes and the CP violating observable
$\epsilon'/ \epsilon$  \cite{Bai:2015nea,Blum:2015ywa,Ishizuka:2018qbn}. In particular, 
a  large enhancement of the $I = 0$ amplitude over the $I = 2$ one has been reported, albeit with too large uncertainty to be considered a
satisfactory first-principles determination of the $\Delta I=1/2$ rule \footnote{While this paper was under revision, a significantly improved result at the physical point was made public \cite{Abbott:2020hxn}.}.

In Ref.~\cite{Boyle:2012ys} an analysis of the different contributions was made and it was suggested that the main source of the enhancement  lies in a strong cancellation of the isospin-two amplitude,  as a result of 
a negative relative sign between the colour-connected and colour-disconnected contractions, with the two contributions adding up in the isospin-zero channel. 
In Refs.~\cite{Donini:2016lwz,Donini:2017rzi,Romero-Lopez:2018rzy} we proposed to study the $N_c$ dependence of the amplitudes, because the two contributions scale differently in large $N_c$ and therefore can be rigorously disentangled in this limit. The enhancement, if explained in this fashion, seems to require unnaturally large-$N_c$ corrections with the appropriate sign. 

Interestingly, the large-$N_c$ limit of QCD \cite{tHooft:1973alw} has also inspired several phenomenological  determinations of these and related observables 
\cite{Bardeen:1986vz,Bardeen:1986vp, Bardeen:1986uz,Chivukula:1986du,Sharpe:1987cx,Pich:1995qp,Hambye:2003cy,Aebischer:2018rrz,Buras:2018lgu} (for a recent discussion see \cite{Buras:2014maa,Gisbert:2018tuf,Cirigliano:2019cpi}). 
It is well known, however, that the leading-order large-$N_c$ prediction for the ratio of the amplitudes, $\lim_{N_c \to \infty} A_0/A_2 = \sqrt{2}$, 
i.e., no $\Delta I=1/2$ rule whatsoever. 
The subleading $N_c$ corrections should therefore be very large, which could be consistent with the previous hypothesis, but casts doubts on the phenomenological approaches that make use of large-$N_c$ inspired approximations: if we know that 
there must be significant large-$N_c$ corrections to explain the $\Delta I=1/2$, why should we trust approximations that neglect subleading $N_c$ terms?
  
The $N_c$ dependence can be studied from first-principles in lattice QCD by simply simulating at different number of colours \cite{Lucini:2001ej,Bali:2013kia,DeGrand:2016pur,Romero-Lopez:2019gqt,Hernandez:2019qed}. In our previous work \cite{Donini:2016lwz,Donini:2017rzi,Romero-Lopez:2018rzy} we explored the related weak amplitudes $K \rightarrow \pi$ and $K \rightarrow \bar{K}$  in the quenched approximation, and found no unnaturally large subleading $N_c$ corrections, although we confirmed
 the exact anticorrelation of these corrections in the two isospin channels. The quenched approximation introduces however an uncontrollable systematic error, which 
 in practice is often found to be relatively small in most quantities. Since we are interested in subleading $N_c$ corrections, quenching effects are expected to enter
 at this order of the $N_c$ expansion and therefore need to be included. The main goal of this paper is to extend our previous study beyond the quenched approximation, which will 
 allow us to determine from first-principles the subleading $N_c$ corrections  to the $\Delta I=1/2$ rule, in a simplified setting with four degenerate flavours, $m_u=m_d=m_s=m_c$. 

This paper is organized as follows:
in Section~\ref{sec:strategy} we discuss our strategy for the lattice study of $K\to \pi$ transitions;
in Section~\ref{sec:scaling} we discuss the $N_c$ scaling of the amplitudes;
Section~\ref{sec:chiral} deals with the necessary results in Chiral Perturbation Theory 
to connect to $K\to\pi\pi$;
Section~\ref{sec:lattice} describes the setup of our lattice computations;
in Section~\ref{sec:results} we discuss our physics results;
and  we conclude in Section~\ref{sec:conclusions}.

\section{Strategy }
\label{sec:strategy}

The Operator Product Expansion  allows to represent CP-conserving $\Delta S=1$ transitions by an effective Hamiltonian of four-fermion operators. At the electroweak scale, $\mu \simeq M_W$, we can neglect all quark masses, and the weak Hamiltonian takes the simple form:
\begin{gather}
\label{eq:heffs1}
H_{\rm w}^{\Delta S=1} = \int d^4x~\frac{g_{\rm w}^2}{4M_W^2}V_{us}^*V_{ud}\sum_{\sigma=\pm} k^\sigma(\mu) \, \bar{Q}^\sigma(x,\mu)\,, 
\end{gather}
where $g_{\rm w}^2=4\sqrt{2}G_{\rm F} M_W^2$. Only two four-quark operators of dimension six can appear with the correct symmetry properties under the flavour 
symmetry group  ${\rm SU}(4)_{\rm L} \times {\rm SU}(4)_{\rm R}$, namely
\begin{gather}
\begin{split}
{\bar Q}^\pm(x,\mu) = Z_Q^\pm(\mu) \, \big(& J_\mu^{su}(x)J_\mu^{ud}(x) \pm J_\mu^{sd}(x)J_\mu^{uu}(x) \\
&~-~[u\leftrightarrow c]\big)\,,
\end{split}
\end{gather}
where $J_\mu$ is the left-handed current $J_\mu^{ij} = (\bar\psi^i\gamma_\mu P_-\psi^j)$; $i,j$ are quark flavour indices; 
$P_\pm={1\over 2} (\mathbf{1}\pm\gamma_5)$; and parentheses around quark bilinears indicate that they are traced\footnote{This basis can be related to the more traditional one by means of Fierz identities.} over spin and colour. 
$Z_Q^\pm (\mu)$ is the renormalization constant of the bare operator $Q^\pm (x)$ computed in some regularization scheme as, for example, the lattice.
There are other operators that could mix with those above: however, they vanish in the limit of equal up and charm masses, that we refer to as the GIM limit \cite{Giusti:2004an}.  From the lattice point of view the GIM limit is very advantageous, not only for the simpler operator mixing, but also because no closed quark propagator contributes to the amplitudes. Even though the presence of a heavy charm was  argued  long ago to be at the origin of the $\Delta I=1/2$ rule via the mixing with 
penguin operators \cite{Shifman:1975tn}, the relevance of penguin contributions has been found to be small in non-perturbative studies \cite{Bai:2015nea,Endress:2014ppa}.\footnote{The dominance of current-current operators over penguin contributions was also pointed out in the Dual QCD approach \cite{Bardeen:1986vz}.} If we want to test the primary mechanism of the $\Delta I=1/2$ enhancement proposed in \cite{Boyle:2012ys}, the GIM limit may be good enough. 

The operators ${\bar Q}^\sigma(\mu)$ are renormalized at a scale $\mu$ in some renormalization scheme, being their $\mu$ dependence exactly cancelled by that of the Wilson coefficients $k^\sigma(\mu)$. It is also possible to define renormalization group invariant (RGI) operators, which are defined by cancelling their  $\mu$ dependence, as derived from the Callan-Symanzik equations,
\begin{eqnarray}
\hat{Q}^\sigma \equiv \hat{c}^\sigma(\mu) {\bar Q}^\sigma(\mu), 
\end{eqnarray}
with 
\begin{eqnarray}
{\hat c}^\sigma(\mu)\equiv &&\left({N_c \over 3} {\bar{g}^2(\mu)\over 4 \pi} \right)^{{\gamma^\sigma_0\over 2 b_0}} \times \nonumber \\&&
\exp\left\{ -\int_0^{\bar{g}(\mu)}{\rm d}g
\left[{\gamma^\sigma(g)\over \beta(g)} - {\gamma^\sigma_0\over b_0 \, g}\right]\right\}, 
\label{eq:c}
\end{eqnarray}
where $\bar{g}(\mu)$ is the running coupling and $\beta(g)=-g^3 \sum_n b_n g^{2 n}$, $\gamma^\sigma(g) = -g^2 \sum_n \gamma^\sigma_n g^{2 n}$ are the $\beta$-function and the four-fermion operator anomalous dimension, respectively. The one- and two-loop coefficients of the $\beta$-function, and the one-loop coefficient
of the anomalous dimensions, are renormalization scheme-independent. Their values for the theory with $N_f$ flavours are \cite{b0,b1_a,b1_b,b1_c}
\begin{eqnarray}
b_0 &=& \frac{1}{(4\pi)^2}\left[\frac{11}{3}N_c-\frac{2}{3}N_f\right]\,,\\
b_1 &=& \frac{1}{(4\pi)^4}\left[\frac{34}{3}N_c^2-\left(\frac{13}{3}N_c-\frac{1}{N_c}\right)N_f\right]\,,
\end{eqnarray}
and for the operators $Q^\pm$ \cite{Gaillard:1974nj,Altarelli:1974exa}
\begin{gather}
\gamma_0^\pm = \frac{1}{(4\pi)^2}\left[\pm 6 - \frac{6}{N_c}\right]\,.
\end{gather}
The normalization of $\hat{c}^\sigma(\mu)$ coincides with the most popular one for $N_c=3$, whilst using the 't Hooft coupling $\lambda = N_c \bar{g}^2 (\mu)$
in the first factor instead of the usual coupling, so that the large-$N_c$ limit  is well-defined.  

Defining similarly an RGI Wilson coefficient 
\begin{eqnarray}
\hat{k}^\sigma \equiv {k^\sigma(\mu)\over \hat{c}^\sigma(\mu)},
\end{eqnarray}
we can rewrite the Hamiltonian in terms of RGI quantities, which no longer depend on the scale, so that
\begin{gather}
\begin{split}
\hat{k}^\sigma \, \hat{Q}^\sigma  &= \left[ { k^\sigma(M_W) \over \hat{c}^\sigma(M_W) } \right] \, \left[ {\hat c}^\sigma(\mu)\, \bar{Q}^\sigma (\mu) \right]\\ 
&= k^\sigma(M_W) \, U^\sigma(\mu,M_W) \, \bar{Q}^\sigma (\mu) \, ,
\end{split}
\end{gather}
where $\mu$ is a convenient renormalization scale for the non-perturbative computation of matrix elements of $Q^\pm$, which will be later set to the inverse lattice scale $a^{-1}$. The factor $U^\sigma(\mu, M_W) = {\hat c}^\sigma(\mu)/ {\hat c}^\sigma(M_W)$, therefore, measures the running of the renormalized operator between the scales $\mu$ and $M_W$.
Ideally one would like to evaluate this factor non-perturbatively, as has been done for $N_c=3$ \cite{Dimopoulos:2007ht,Guagnelli:2005zc}, but such a challenging endeavour is beyond the scope of this paper. We will instead use 
the perturbative results at two loops in the RI scheme \cite{Ciuchini:1997bw,Buras:2000if}
to evaluate the Wilson coefficients $k^\sigma(M_W)$, the running factors $U^\sigma(\mu,M_W)$, and $\hat c(\mu)$. This implies relying on perturbation theory at scales
above $\mu\geq a^{-1} \sim 2.6~{\rm GeV}$. Similarly we will also use lattice perturbation theory to estimate the renormalization factors $Z_Q^\pm$, that are known to one loop\footnote{The NLO running of the coupling and four-quark operators have been performed fully in the $N_f=4$
theory, using the value of $\Lambda_{\overline{\rm MS}}(N_f=4)$ by the ALPHA Collaboration
in Ref.~[42]. We have checked that the effect of running from $N_f=5$ from $M_W$
to the $b$ quark mass, and then with $N_f=4$ down to the lattice matching scale amounts to few per mille effects on the running factors. This is completely negligible within the uncertainty of our final results.}  \cite{Alexandrou:2010me,Alexandrou:2012mt}. 

We are interested in considering $K\to \pi$ amplitudes in the two isospin channels, that we can extract from ratios of three-point correlators
\begin{align}
\begin{split}
{\mathcal C}_3^\pm(y,&\,z,x) \equiv \\ 
&  \langle P^{du}(y) [O^{suud}(z)\pm O^{sduu}(z)] P^{us}(x) \rangle,
\label{eq:pqp}
\end{split}
\end{align}
where 
\begin{eqnarray}
P^{ij}(x) \equiv \bar{\psi}^i(x)\gamma_5 \psi^j(x),\;\;O^{ijkl} \equiv \bar{\psi}^i \gamma_\mu \psi^j \bar{\psi}^k \gamma_\mu \psi^l,
\end{eqnarray}
and the two-point correlators
\begin{eqnarray}
{\mathcal C}^{ij}_{2}(y,z) &\equiv& \langle P^{ij}(y) A_0^{ji}(z) \rangle,  
\label{eq:2pt} 
\end{eqnarray}
with $A_0^{ij}(x) \equiv \bar{\psi}^i(x)\gamma_0 \gamma_5 \psi^j(x)$. 

From these correlators we define the bare lattice ratios:
\begin{eqnarray}
R^\pm  = \kern-1.0em
\lim_{ \substack{z_0-x_0\to\infty \\ y_0-z_0\to \infty}}
\frac{\sum_{{\mathbf x},{\mathbf y}}{\mathcal C}_3^\pm(y,z,x)}
{\sum_{{\mathbf x,\mathbf y}} {\mathcal C}^{du}_2(y,z) {\mathcal C}^{us}_2(x,z)}, \label{eq:ratios}
\end{eqnarray}
which are proportional to the $K \to \pi$ matrix elements with a convenient normalization.
The renormalization factors for these ratios, $Z^\pm$, are obtained from the ratio of the renormalization factors of the four fermion operators,  and the current normalization factors that appear in the denominator. 

From the renormalized ratios
\begin{equation}
\bar{R}^\sigma = Z^\sigma R^\sigma,
\end{equation}
we can obtain the RGI normalized ratios
\begin{equation}
\hat{R}^\sigma = \hat{c}(a^{-1}) Z^\sigma R^\sigma,
\end{equation}
and the normalized\footnote{Note that our normalization in Eq. \eqref{eq:ratios} cancels two powers of the decay constant in the physical amplitudes.} $K\rightarrow \pi$ amplitudes, written either in terms of the RGI or the renormalized ratios, as 
\begin{equation}
A^\sigma = \hat{k}^\sigma \hat{R}^\sigma = k^\sigma(M_W) U^\sigma(a^{-1},M_W) \bar{R}^\sigma.
\label{eq:apm}
\end{equation}
All the required factors to reconstruct the physical amplitudes are summarized in Table~\ref{tab:renorm} for $N_f=4$ (this work), and in Table~\ref{tab:renormquenched} for the quenched case~\cite{Donini:2016lwz,Donini:2017rzi}. 
\begin{table}[!t]
\begin{center}
\begin{tabular}{c@{\hspace{5mm}}ccccc}
\hline\hline\\[-2.0ex]
$N_c$ &
$k^+(M_W)$ & $U^+(a^{-1},M_W)$ &  $Z^+(a^{-1})$ & $\hat{c}^+(a^{-1})$\\[0.3ex] 
\hline\\[-2.0ex]
 3       & 1.041 & 0.843 & 0.841 & 1.456 \\
 \ $3^*$ & 1.041 & 0.852 & 0.844 & 1.471 \\
 4       & 1.032 & 0.877 & 0.884 & 1.367 \\
 5       & 1.026 & 0.899 & 0.909 & 1.302 \\
 6       & 1.022 & 0.914 & 0.926 & 1.255 \\
 \hline\hline 
\end{tabular}
\vspace{5mm}

\begin{tabular}{c@{\hspace{5mm}}cccc}
\hline\hline\\[-2.0ex]
$N_c$ &
$k^-(M_W)$ & $U^-(a^{-1},M_W)$  & $Z^-(a^{-1})$ &  $\hat{c}^-(a^{-1})$\\[0.3ex] 
\hline\\[-2.0ex]
  3       & 0.918 & 1.433 & 1.320 & 0.488 \\
 \ $3^*$ & 0.918 & 1.400 & 1.314 & 0.476 \\
 4       & 0.947 & 1.254 & 1.195 & 0.602 \\
 5       & 0.961 & 1.179 & 1.137 & 0.679 \\
 6       & 0.970 & 1.137 & 1.104 & 0.731 \\
 \hline\hline
\end{tabular}
\caption{Perturbative renormalization constants and RG running factors for the ensembles with $N_f=4$. $Z^\sigma (a^{-1})$
have been computed at one loop in tadpole-improved perturbation theory using the results in~\cite{Alexandrou:2010me,Alexandrou:2012mt}, 
whereas $U^\sigma$ and $k^\sigma$ are computed using the two-loop $\overline{\rm MS}$ coupling. The star labels the simulation points with finer lattice spacing, $a \sim 0.065 $ fm. In the evaluation of  $\hat{c}^\sigma(a^{-1})$ we have used $\Lambda_{\overline{\text{MS}}}(N_f=4)= 298 $ MeV from Ref. \cite{Bruno:2017gxd}. }
\label{tab:renorm} 
\end{center}
\end{table}

\begin{table}[!t]
\begin{center}
\begin{tabular}{c@{\hspace{5mm}}ccccc}
\hline\hline\\[-2.0ex]
$N_c$ &
$k^+(M_W)$ & $U^+(a^{-1},M_W)$ &  $Z^+(a^{-1})$ & $\hat{c}^+(a^{-1})$\\[0.3ex] 
\hline\\[-2.0ex]
 3 & 1.029 & 0.877 &  0.956 & 1.412\\
 4 & 1.025 & 0.897 &  0.963 & 1.340\\
 5 & 1.021 & 0.911 &  0.969 & 1.285\\
 6 & 1.018 & 0.923 &  0.973 & 1.243\\
 7 & 1.016 & 0.932 &  0.976 & 1.212\\
 8 & 1.014 & 0.939 &  0.979 & 1.187\\
 17& 1.007 & 0.969 &  0.989 & 1.091
\\ \hline\hline 
\end{tabular}
\vspace{5mm}

\begin{tabular}{c@{\hspace{5mm}}cccc}
\hline\hline\\[-2.0ex]
$N_c$ &
$k^-(M_W)$ & $U^-(a^{-1},M_W)$  & $Z^-(a^{-1})$ &  $\hat{c}^-(a^{-1})$\\[0.3ex] 
\hline\\[-2.0ex]
 3 & 0.942 & 1.312 &  1.087 & 0.511 \\
 4 & 0.959 & 1.206 &  1.061 & 0.619 \\
 5 & 0.969 & 1.153 &  1.047 & 0.690 \\
 6 & 0.975 & 1.121 &  1.038 & 0.740 \\
 7 & 0.979 & 1.101 &  1.032 & 0.776 \\
 8 & 0.982 & 1.086 &  1.027 & 0.803 \\
 17& 0.992 & 1.037 &  1.012 & 0.907 \\
 \hline\hline
\end{tabular}
\caption{Perturbative renormalization constants and RG running factors for the runs with $N_f=0$ of Refs.~\cite{Donini:2016lwz,Donini:2017rzi}.  $Z^\sigma (a^{-1})$
have been computed at one loop in tadpole-improved perturbation theory using the results in~\cite{Alexandrou:2010me,Alexandrou:2012mt}, whereas $U^\sigma$ and $k^\sigma$ are computed using the two-loop $\overline{\rm MS}$ coupling.
Note that the values of $Z^\sigma (a^{-1})$ differ from those in Refs.~\cite{Donini:2016lwz,Donini:2017rzi}, where bare lattice perturbation theory was used. Furthermore, the values of $k^\sigma$ and $U^\sigma$ also supersede the ones in Refs.~\cite{Donini:2016lwz,Donini:2017rzi}. In the evaluation of  $\hat{c}^\sigma(a^{-1})$ we have used $\Lambda_{\overline{\text{MS}}}$ as described in Ref. \cite{Donini:2016lwz}.
}
\label{tab:renormquenched} 
\end{center}
\end{table}

\section{large-$N_c$ scaling of $K\rightarrow \pi$ amplitudes }
\label{sec:scaling}
\subsection{Diagrammatic expansion of $A^\pm$ }
A simple diagrammatic analysis of the three and two point correlators of Eqs.~(\ref{eq:pqp},\ref{eq:2pt}) shows a clear pattern of the large-$N_c$ scaling, and demonstrates
the expected anticorrelation of the leading large-$N_c$ corrections  of the $A^\pm$ amplitudes.

\begin{figure}[h!]
 \begin{center}
\renewcommand{\arraystretch}{2}
\[\begin{array}{lll}
 \raisebox{-.5\height}{\includegraphics[scale=0.7]{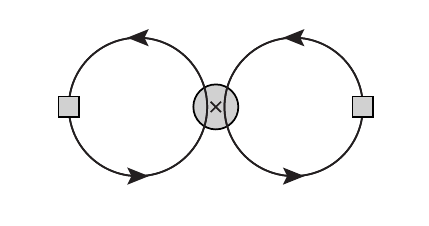}} & \mp &\raisebox{-.5\height}{\includegraphics[scale=0.7]{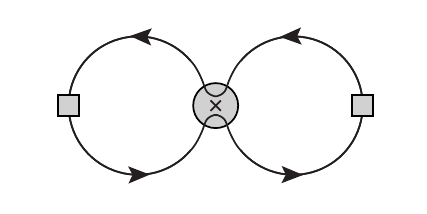}}\\
 \end{array}\]
 \caption{\label{fig:condis} Left diagram:  $O^{suud}(x)$ insertion or colour-disconnected contribution to ${\cal C}^\pm_3$ in Eq.~(\ref{eq:pqp}).
 Right diagram: $O^{sduu}(x)$ insertion or colour-connected contribution to ${\cal C}^\pm_3$ in Eq.~(\ref{eq:pqp}). }
\end{center}
\end{figure}

\begin{figure}[h]
 \begin{center}
\renewcommand{\arraystretch}{2}
\[\begin{array}{ccc}
(a) & \raisebox{-.5\height}{\includegraphics[scale=0.7]{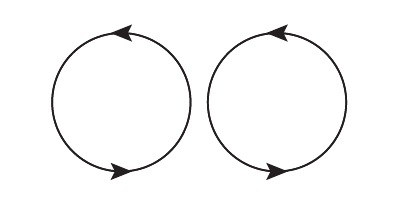}} &  {\mathcal O}(N_c^2) \\
(b) & \raisebox{-.5\height}{\includegraphics[scale=0.7]{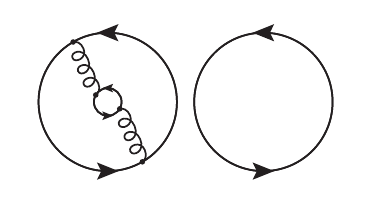}} & {\mathcal O}(N_c N_f)  \\
(c) & \raisebox{-.5\height}{\includegraphics[scale=0.7]{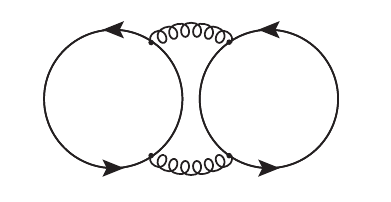}} &{\mathcal O}(N_c^0) \\
(d) & \raisebox{-.5\height}{\includegraphics[scale=0.7]{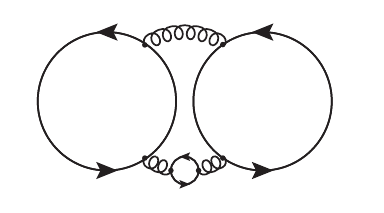}} & {\mathcal O}\left({N_f\over N_c}\right)\\
\end{array}\]
\caption{\label{fig:disnlo} $N_c, N_f$ scaling of various contributions to the colour-disconnected contraction, corresponding to the $O^{suud}(x)$ insertion.}
\renewcommand{\arraystretch}{2}
\[ \begin{array}{cc}
  (e)   & \left \{ 
\begin{array}{cc}
\raisebox{-.5\height}{\includegraphics[scale=0.7]{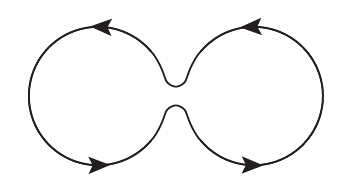}} & {\mathcal O}(N_c)  \\
\raisebox{-.5\height}{\includegraphics[scale=0.7]{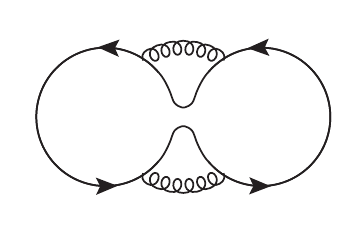}} & {\mathcal O}(N_c)
\end{array} 
  \right . \\
(f) &  \hspace{0.405cm} \raisebox{-.5\height}{\includegraphics[scale=0.7]{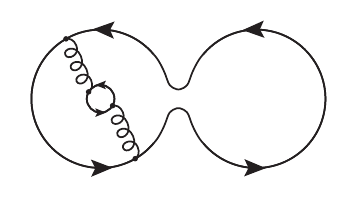}}  \hspace{0.23cm} {\mathcal O}\left({N_f}\right)\\
\end{array}\]
\caption{\label{fig:connlo} $N_c, N_f$ scaling of various contributions to the colour-connected contraction, corresponding to the $O^{sduu}(x)$ insertion. }
\end{center}
\end{figure}

After integration over fermion fields, the correlators are obtained from the gauge averages of the colour-disconnected and colour-connected contractions of Fig.~\ref{fig:condis}, 
corresponding to the operator insertion $O^{suud}$ and $O^{sduu}$, respectively.
In Figs.~\ref{fig:disnlo} and~\ref{fig:connlo} we show the scaling with $N_c$ of the lowest-order diagrams contributing to these correlators. 
The leading $N_c$ dependence of both the renormalized and bare correlators 
are therefore of the form:
\begin{eqnarray}
\langle P^{ij} J_\mu^{ji} \rangle  & =& N_c \left(a + b {N_f\over N_c}\right)+\ldots, \nonumber\\
\langle P^{du}  O^{suud} P^{us} \rangle &=& \langle P^{du} J_\mu^{ud} \rangle  \langle P^{su} J_\mu^{us} \rangle+ c+ d {N_f\over N_c}  + \ldots ,\nonumber\\
\langle P^{du}  O^{sduu} P^{us} \rangle &=& N_c\left(e + f {N_f\over N_c}\right)+ \ldots ,
\label{eq:largenc}
\end{eqnarray}

\noindent where all the coefficients $a-f$ in these expressions (each of them related to one or more diagrams in Figs.~\ref{fig:disnlo} and~\ref{fig:connlo})
are independent of $N_c$ and $N_f$. These relations imply that 
the leading $\NC$ corrections in the $\pm$ correlation functions of Eq.~(\ref{eq:pqp}) are of ${\mathcal O}(N_c^2, N_f N_c)$, but factorizable.  
On the other hand, the leading non-factorizable corrections are of ${\mathcal O}(N_c)$ and ${\mathcal O}(N_f)$, and cancel in the sum of the 
 $\pm$ correlators:
 \begin{eqnarray}
{\mathcal C}_3^+ + {\mathcal C}_3^-&=& {\rm disconnected}+ {\mathcal O}(N_c^0) + {\mathcal O}\left({N_f\over N_c}\right)  + \cdots ,\nonumber\\
{\mathcal C}_3^+ - {\mathcal C}_3^- &=& {\mathcal O}(N_c) + {\mathcal O}(N_f) + \cdots  \label{eq:CplusCminus}
 \end{eqnarray}
  They are therefore fully anticorrelated in the $\pm$ correlators. Importantly, the anticorrelated terms include the leading 
fermion loop corrections, ${\mathcal O}(N_f)$.  These relations also imply the following scaling of the renormalization factors:
\begin{eqnarray}
{Z_Q^++Z_Q^-\over 2} &=& 1+  {\mathcal O}\left({1\over N_c^2}\right) + {\mathcal O}\left({N_f\over N_c^3}\right)+\cdots\nonumber\\
{Z_Q^+-Z_Q^-\over 2} &=& {\mathcal O}\left({1\over N_c}\right) + {\mathcal O}\left({N_f \over N_c^2} \right)+\cdots,
\end{eqnarray}
and a similar one for the Wilson coefficients, $k^\sigma$.  
This dependence can be explicitly checked in the perturbative coefficients known up to two loops in the $\overline{\rm MS}$ scheme \cite{Ciuchini:1997bw,Buras:2000if}.

These results imply the following scaling of the amplitudes:
\begin{equation}
A^\pm = 1 \pm \tilde a {1 \over N_c}\pm \tilde b {N_f \over N^2_c}+\tilde c {1 \over N^2_c}+ \tilde d {N_f \over N^3_c}+\cdots, \label{eq:rnc}
\end{equation}
where the coefficients $\tilde a- \tilde d$ are combinations of the coefficients $a-f$ in  Eq.~(\ref{eq:largenc}), and are also
independent of $N_c$ and $N_f$, and  a {\it natural} expectation is that they are ${\mathcal O}(1)$.

Not only the leading corrections $N_c^{-1}$ are, therefore, fully anticorrelated in the ratios, 
but also the leading effects of dynamical quarks, ${\mathcal O}(N_f)$. Note that this analysis 
does not predict the sign of the  different terms, i.e., the sign of the $\tilde a-\tilde d$ coefficients, only the (anti)-correlation between the two isospin channels. This way, a negative sign of $\tilde a$ and $\tilde b$ results into an enhancement of the ratio $A^-/A^+$.

\subsection{ 't Hooft vs. Veneziano scaling}

As we will see the number of active flavours, $N_f$, plays a relevant role in the $1/N_c$ expansion of the $K \to \pi$ amplitudes. The scaling in $N_f$ is in fact the difference between the 't Hooft and Veneziano limits of QCD. While the former keeps $N_f$ constant when taking $N_c \to\infty $, the latter keeps the ratio $N_f/N_c$ constant. From Eq.~(\ref{eq:rnc}), it is then clear that $\tilde{a}$ and $\tilde{b}$ have the same scaling in the Veneziano limit (the same holds for $\tilde{c}$ and $\tilde{d}$). In our simulations, we will be studying the 't Hooft limit, since we keep $N_f$ fixed, but the quantity $N_f/N_c$ is large (ranging from 4/3 to 2/3, depending on $N_c$), so its contribution may be very significant even for {\it naturally} large $\tilde{a}-\tilde{d}$ coefficients. 

\section{$\Delta S=1$ amplitudes in Chiral Perturbation Theory}
\label{sec:chiral}

\subsection{ Chiral Dependence of the $K \to \pi$ amplitudes }

The chiral dependence of the ratios in Eq.(\ref{eq:ratios}) can be studied within the framework of Chiral Perturbation Theory (ChPT) with $N_f=4$ active flavours. An extensive discussion of this framework can be found in Refs.~\cite{Giusti:2004an,Hernandez:2006kz}. Here we just summarize the required formul{\ae}, and refer to those references for details.

The weak Hamiltonian in Eq. (\ref{eq:heffs1}) can be translated to an effective weak Hamiltonian in terms of meson fields preserving the flavour symmetries. Since the operators $\bar{Q}^+$ and $\bar{Q}^-$ transform under representations of $SU(4)_L$ of dimension {\bf 84} and {\bf 20}, their ChPT counterparts must be constructed accordingly. At leading order, there are only two terms, with couplings $g^\pm$, that need to be determined non-perturbatively:
\begin{equation}
\mathcal{H}_W^{ChPT} = g^+ \mathcal{O}^+ + g^- \mathcal{O}^-, \label{eq:weakHchpt}
\end{equation}
with
\begin{equation}
\mathcal{O}^\sigma = \sum_{ijkl} c^\sigma_{ijkl} F^4 (U \partial_\mu U^\dagger)_{ij} (U \partial^\mu U^\dagger)_{kl},
\end{equation}
where $U$ is the chiral meson field, $i,j,k,l$ are flavour indices, and $c^\sigma_{ijkl}$ are Clebsch-Gordan coefficients (see Appendix A in Ref.~\cite{Giusti:2004an}).

By means of the chiral weak Hamiltonian in Eq.~(\ref{eq:weakHchpt}) and the standard NLO ChPT Lagrangian, the chiral predictions for the normalized amplitudes in Eq.~(\ref{eq:apm}) are found to be:\begin{equation}
A^\pm = g^\pm \left[  1\mp 3 \left( \frac{M_\pi}{4 \pi F_\pi} \right)^2 \left( \log \frac{M_\pi^2}{\mu^2} + L_\pm^r(\mu) \right)    \right], \label{eq:ratioschiral}
\end{equation}
where $L^r_\pm$ are the NLO counterterms\footnote{$L^r_\pm$ are a combination of standard QCD NLO LECs with those associated to higher order operators in the chiral weak Hamiltonean. See Refs. \cite{Kambor:1989tz} and \cite{Hernandez:2006kz} for explicit expressions.}. The NLO corrections in Eq.~(\ref{eq:ratioschiral}) are fully anticorrelated.  Extrapolating the ratios in Eq.~(\ref{eq:ratios}) to zero pion mass, one can determine the leading low-energy couplings (LECs) of the chiral weak Hamiltonian:
\begin{equation}
g^\pm = \lim_{M_\pi \to 0} A^\pm.
\end{equation}
The extracted values of $g^\pm$ can then be used to make predictions of other observables, such as the $K \to \pi \pi$ decay amplitudes.

We now turn to the analysis of the combined chiral and $N_c$ dependence. First, we note that Eq.~(\ref{eq:rnc}) should hold at any pion mass, and therefore we expect:
\begin{align}
\begin{split}
g^\pm = 1 &\pm  a_\chi {1 \over N_c} \pm  b_\chi {N_f \over N^2_c} +  c_\chi {1 \over N^2_c}+  d_\chi {N_f \over N^3_c}  + \cdots \label{eq:gNc}
\end{split}
\end{align}
Furthermore, by comparing the chiral dependence in Eq.~(\ref{eq:ratioschiral}) with the $N_c$ scaling in Eq.~(\ref{eq:rnc}) we can see that both $L^r_+$ and $L_-^r$ must be $O(N_c^0)$, and identical at this order. The next term in the $1/N_c$ expansion for $L^r_\pm$ could in principle differ:
\begin{equation}
L^r_\pm = L^{(0)} + \frac{1}{N_c} L^{(1)}_{\pm}  + \cdots. \label{eq:LNc}
\end{equation}
Hence, the combination of Eq.~(\ref{eq:ratioschiral}) with Eqs.~(\ref{eq:gNc},\ref{eq:LNc}) can be used to do global fits including different meson masses and values of $N_c$.

It will be convenient to also study the chiral and $N_c$ dependence of the product of $A^+ A^-$. The reason is that the leading chiral and $N_c$ corrections cancel out, which leads to a more robust chiral extrapolation. The chiral corrections for this quantity are
\begin{align}
\begin{split}
A^+ A^- &= g^+ g^- \left[1 +3 \left( \frac{M_\pi}{4 \pi F_\pi} \right)^2 {(  L^r_- - L^r_+ )} \right], \label{eq:fitprod}
\end{split}
\end{align}
with
\begin{eqnarray}
g^+ g^- &=& 1 + \alpha {1 \over N_c^2} + \beta {1 \over N_c^3}+ \hdots, \\
L^r_- - L^r_+ &=&  \frac{  L^{(1)}_- - L^{(1)}_+ }{N_c} + \hdots, \label{eq:Ncprod}
\end{eqnarray}
where $\alpha$ and $\beta$ depend on the coefficients $a_\chi - d_\chi$.

\subsection{Relation to $K \to \pi \pi$ amplitudes}

Once the effective couplings $g^\pm$ have been extracted from the chiral extrapolations of the ratios $A^\pm$, they can be used to compute the $K \to \pi \pi$ weak decay amplitudes. The two pions in the final state can be in a state with total isospin $I=0$ or $2$:
\begin{equation}
iA_Ie^{i\delta_I} = \braket{\left( \pi \pi \right)_I | \mathcal{H}_W^{ChPT} |K^0 },
\end{equation}
where $\delta_I$ is the two-pion scattering phase.
The ratio of the two amplitudes can be calculated at leading order in ChPT using the Hamiltonian in Eq.~(\ref{eq:weakHchpt}) \cite{Giusti:2004an,Giusti:2006mh}:
\begin{equation}
\frac{A_0}{A_2} = \frac{1}{2\sqrt{2}} \left( 1 + 3 \, \frac{g^-}{g^+} \right).
\end{equation}
The measured hierarchy of $\sim 22$ between $A_0$ and $A_2$ must then be translated into a large ratio of the couplings $g^\pm$. Note that for $g^+ = g^-  = 1$, the expected large-$N_c$ result is recovered,  $A_0/A_2 = \sqrt{2}$.  Large $1/N_c$ corrections in the $g^-/g^+$ ratio could therefore be the origin of the $\Delta I=1/2$ rule.

We have also derived the ChPT NLO result for the non-degenerate case in which  we send the pion mass to zero, while keeping the kaon mass at its physical value\footnote{See Ref. \cite{Golterman:1997wb} for similar calculation in $N_f=3$ ChPT.}. As we are forced to work in the exact GIM limit, we must also send the charm quark mass to zero with the up quark mass. The calculation for $m_s > m_u = m_d = m_c =0$ yields:
\begin{align}
\begin{split}
 \text{Re } &\frac{A_0}{A_2} \Big \rvert_{M_\pi,M_D \to0, M^{\text{phys}}_K}= {\frac{1}{2\sqrt{2}} \left( 1 + 3 \, \frac{g^-}{g^+} \right)  }\\ & {+ \frac{17}{12\sqrt{2}}  \left( 1+ \frac{1}{17}\frac{g^-}{g^+} \right) \frac{M_K^2}{(4 \pi F_K)^2} \log \frac{\Lambda_{\rm eff}^2}{M_K^2}} \, ,
 \label{eq:NLOktopipi}
\end{split}
\end{align}
where $\Lambda_{\rm eff}$ is an unknown scale that contains information of the NLO LECs of the effective Chiral Lagrangian and the effective weak Hamiltonian.  We note that the NLO effect tends to enhance (reduce) the ratio for $\Lambda_{\rm eff} > M_K$ ($\Lambda_{\rm eff} < M_K$).

\section{Lattice setup}
\label{sec:lattice}
 
\begin{table}[h!]
\centering
\begin{tabular}{c|c|c|c|c|c|c}
Ensemble & $\NC$& $\beta$& $c_{\rm\scriptscriptstyle sw}$& $T \times L$   & $am^\text{s}_0$ & \# configs  \\ \hline \hline
3A10 &\multirow{ 5}{*}{3}&\multirow{ 5}{*}{1.778} & \multirow{ 5}{*}{1.69}& $36 \times 20$ & -0.4040  & 195 \\  \cline{1-1}  \cline{5-7} 
3A11 && & &$48 \times 24$ & -0.4040 & 81  \\  \cline{1-1}  \cline{5-7} 
3A20 & & & &$48 \times 24$& -0.4060 & 155\\  \cline{1-1}  \cline{5-7} 
3A30 & & & &$48 \times 24$& -0.4070 & 149 \\  \cline{1-1}  \cline{5-7} 
3A40 & & & &$60 \times 32$& -0.4080 & 94 \\ \hline   \hline 
3B10 & \multirow{ 2}{*}{3}  & \multirow{ 2}{*}{1.820}&\multirow{ 2}{*}{1.66} & $48 \times 24$& -0.3915 & 182 \\  \cline{1-1}  \cline{5-7} 
3B20 &  &  && $60 \times 32$& -0.3946 & 164 \\ \hline   \hline 
4A10 &\multirow{ 3}{*}{4}  & \multirow{ 3}{*}{3.570}& \multirow{ 3}{*}{1.69}& $36 \times 20$& -0.3725 & 82 \\    \cline{1-1}  \cline{5-7} 
4A30 & & & &$48 \times 24$& -0.3760 & 153\\ \cline{1-1}  \cline{5-7} 
4A40 & & & &$60 \times 32$& -0.3780 & 55 \\ \hline \hline
5A10 &\multirow{ 3}{*}{5}  & \multirow{ 3}{*}{5.969}&\multirow{ 3}{*}{1.69}& $36 \times 20$& -0.3458  & 52\\   \cline{1-1}  \cline{5-7} 
5A30 & & & &$48 \times 24$& -0.3500 & 39 \\ \cline{1-1}  \cline{5-7} 
5A40 & & & &$60 \times 32$& -0.3530 & 36 \\ \hline  \hline
6A10 &\multirow{ 3}{*}{6}  & \multirow{ 3}{*}{8.974}&\multirow{ 3}{*}{1.69}& $36 \times 20$& -0.3260 & 35 \\    \cline{1-1}  \cline{5-7} 
6A30 & & & & $48 \times24$& -0.3311 & 30 \\    \cline{1-1}  \cline{5-7} 
6A40 & & & &$60 \times 32$& -0.3340 & 40 \\ \hline 
\end{tabular}
\caption{  Summary of the simulation parameters of the various ensembles used in this work.  }
\label{tab:paramensembles}
\end{table} 

\subsection{Simulation and matching of sea and valence sectors}

Our lattice setup  is the same as the one presented in Ref.~\cite{Hernandez:2019qed}, and we refer to it for details on the simulations and scale setting. We use ensembles with $N_f=4$ dynamical fermions for an $SU(N_c)$ gauge theory, with $N_c=3-6$. They have been generated using the HiRep code \cite{DelDebbio:2008zf,Patella:2010dj}. We have chosen the Iwasaki gauge action (following previous experience with 2+1+1 simulations \cite{Alexandrou:2018egz}) and clover Wilson fermions for the sea quarks, with the plaquette-boosted one-loop value of $c_{\rm\scriptscriptstyle sw}$. The simulation parameters are shown in Table~\ref{tab:paramensembles}. We find that a separation of $\geq 10$ units of Montecarlo time produces no autocorrelation in the ratios. The lattice spacing is found to be $a\sim 0.075$ fm for all values of $N_c$ (see also Ref.~\cite{Hernandez:2019qed}). In addition, we have produced two ensembles with a finer lattice spacing, $a \sim 0.065$ fm, to estimate discretization effects.

In order to achieve automatic $O(a)$ improvement\footnote{
As discussed in~\cite{Ugarrio:2018ghf,Bussone:2019mlt}, there are residual $O(a)$ cutoff effects from virtual sea quarks, which are proportional to $am^{\rm s}$ 
and carry coefficients that are $O(\alpha_{\rm\scriptscriptstyle s}^2)$ in perturbation theory. These 
effects are expected to be numerically very small and thus irrelevant for the discussion below. It is also 
worth stressing that using the one-loop value of $c_{\rm\scriptscriptstyle sw}$ will also lead to residual effects of $O(a \, \alpha_{\rm\scriptscriptstyle s}^2)$.
}
~\cite{Frezzotti:2003ni} and avoid the mixing of different-chirality operators for weak decays, we employ maximally twisted valence quarks \cite{Frezzotti:2000nk}, i.e., the mixed-action setup \cite{Bar:2002nr}  previously used in Refs.~\cite{Ugarrio:2018ghf,Bussone:2019mlt}. Working in twisted quark field variables, maximal twist is ensured by tuning the untwisted bare valence mass $m^{\rm v}$ to the critical value for which the valence PCAC mass is zero:
\begin{equation}
\lim_{m^{\rm v} \rightarrow m_{\rm cr}} m^{\rm v}_{\rm pcac} \equiv \lim_{m^{\rm v} \rightarrow m_{\rm cr}} \frac{\partial_0\braket{A_0^{ij}(x) P^{ji}(y)}}{2 \braket{P^{ij}(x) P^{ji}(y)}}= 0.
\end{equation}
The bare twisted mass parameter $\mu_0$ is tuned such that the pion mass in the sea and valence sectors coincide, $M_\pi^{\rm v}= M_\pi^s$.  

Since twisted mass already provides $O(a)$ improvement, the clover improvement parameter $c_{\rm\scriptscriptstyle sw}$ can be chosen to be an arbitrary value in the valence sector. We choose $c_{\rm\scriptscriptstyle sw}=0$ in the valence sector\footnote{
This differs from Ref.~\cite{Hernandez:2019qed}, where we picked $c_{\rm\scriptscriptstyle sw}=1.69$. This value matches the one in the sea sector.
} 
for this work,
our main motivation being that this minimizes the isospin breaking effects coming from the twisted-mass action.
In addition, this will allow for a partial crosscheck of the systematics due to the use of perturbative renormalization
constants, by comparing the latter to the non-perturbative determination in Ref.~\cite{Carrasco:2015pra} for $N_c=3$ (see below).
Finally, we also observe that $c_{\rm\scriptscriptstyle sw}=0$ leads to smaller statistical errors.

In Table~\ref{tab:ensembles} we present our measurements for the ensembles used in this work. We have achieved good tuning to maximal twist, with the PCAC mass being zero within 1 or 2$\sigma$. In addition, the valence and sea pion masses are matched also  within 1 or 2$\sigma$. The bare results for the ratios are also presented in the same table, together with the chiral parameter $\xi = {M_\pi^2}/{(4\pi F_\pi)^2}$, that will be used for the chiral extrapolations.

We conclude the discussion of the simulation setup by mentioning that we will compare the new results with dynamical fermions to the ones in Refs.~\cite{Donini:2016lwz,Donini:2017rzi}. Those results used quenched simulations, with plaquette gauge action and twisted mass fermions. The lattice spacing was $a\sim 0.093$ fm and the the pion mass was fixed at around $M_\pi = 550-590$ MeV for $N_c =3-8$ and $17$. In this work, we perform a reanalysis of these quenched data. 

\begin{table*}[th]
\centering
\begin{tabular}{c|c|c|c|c|c|c|c|c|c|c}
Ensemble & $\NC$&   $aM^\text{s}_\pi$  & $am^\text{tm}_0$ &  $a\mu_0$&  $aM^\text{v}_\pi$ &  $|am^\text{v}_{\text{pcac}}|$  &  $R^+$ & $R^-$ & $\xi$  & $\xi_L$  \\ \hline \hline
3A10 &\multirow{ 5}{*}{3}  &0.2204(21) & -0.9353 & 0.01150&0.2220(19) & 0.0004(4)  & 0.611(17)&      1.418(20) &0.1685(56)&0.1626(56)  \\  \cline{1-1}  \cline{3-11} 
3A11 &  &0.2147(18) & -0.9353 & 0.01150&0.2184(13) & 0.0004(4)  & 0.627(16)&      1.389(18) &0.1520(35)&0.1504(35)  \\  \cline{1-1}  \cline{3-11} 
3A20 &  &0.1845(14) &   -0.9324  & 0.00815  & 0.1833(12)& 0.0002(5)& 0.582(29)&       1.450(33) & 0.1352(39)& 0.1311(39)\\  \cline{1-1}  \cline{3-11} 
3A30 &  &0.1613(16) &  -0.9311  & 0.00660   & 0.1607(15) & 0.0002(3) & 0.511(44) & 1.531(50)    &0.1240(35)&0.1165(35) \\  \cline{1-1}  \cline{3-11} 
3A40 &  &0.1429(12) &   -0.9285  & 0.00534 &0.1413(12) & 0.0002(5)& 0.554(33)   &    1.480(34) & 0.1033(19)& 0.1013(19) \\ \hline   \hline 
3B10 & \multirow{ 2}{*}{3}   &0.1755(15) & -0.8962 & 0.00849  & 0.1761(11) & 0.0001(3) &  0.589(16) & 1.464(19) & 0.1564(40)  & 0.1495(40)  \\  \cline{1-1}  \cline{3-11} 
3B20 &    &0.1191(9) & -0.8919 & 0.00440  & 0.1206(13) & 0.0005(3) &0.489(23)  & 1.533(24) & 0.1017(30)& 0.0958(31)  \\ \hline   \hline 
4A10 &\multirow{ 3}{*}{4}  & 0.2035(14) &  -0.9058 & 0.01055& 0.2043(28)& 0.0010(7)& 0.766(14)&  1.262(17)  & 0.1007(36)& 0.0978(36) \\    \cline{1-1}  \cline{3-11}
4A30 & & 0.1714(8) &   -0.9040 & 0.00797  & 0.1736(12)&  0.0004(3)& 0.699(20)&    1.358(30) &  0.0803(18)&0.0783(18)\\ \cline{1-1}  \cline{3-11} 
4A40 & &0.1397(8) & -0.9030 & 0.00551   & 0.1418(7) & 0.0003(2)  &  0.699(18)    &   1.379(34)  & 0.0612(10)&0.0605(10) \\ \hline \hline
5A10 &\multirow{ 3}{*}{5}  &0.2128(9) &   -0.8783 & 0.01191 & 0.2112(12)&  0.0005(6) & 0.824(8)&      1.201(14)&0.0735(20)&0.0720(20)  \\   \cline{1-1}  \cline{3-11} 
5A30 & & 0.1712(6) &  -0.8768 & 0.00810  & 0.1706(10) & 0.0001(4) & 0.761(17)  & 1.274(27) & 0.0585(11) & 0.0573(11) \\ \cline{1-1}  \cline{3-11} 
5A40 &  &0.1331(7) &  -0.8753  & 0.00517& 0.1338(10) & 0.0001(3) &  0.760(22)  &     1.302(27)    & 0.0407(10) & 0.0403(10) \\ \hline  \hline
6A10 &\multirow{ 3}{*}{6}  &0.2150(7) &  -0.8562 & 0.01280& 0.2136(9)&0.0001(3) & 0.842(9)  &    1.170(9)  &0.0611(9) &0.0601(9) \\    \cline{1-1}  \cline{3-11} 
6A30 & &0.1689(7) &  -0.8548  &0.00803 &0.1669(7) & 0.0004(3) & 0.821(12) & 1.185(18) & 0.0455(7) & 0.0447(7) \\    \cline{1-1}  \cline{3-11} 
6A40 & &0.1351(6) &  -0.8548& 0.00542& 0.1352(3) & 0.0000(2) & 0.805(9) &   1.219(8)   & 0.0328(3) & 0.0325(3) \\ \hline 
\end{tabular}
\caption{  Summary of results for our ensembles with Iwasaki gauge action and $O(a)$-improved Wilson fermions with $c_{\rm\scriptscriptstyle sw}=0$ in the valence sector throughout. The value of the lattice spacing is $a \simeq 0.075\text{ fm}$ for the ``A'' ensembles (see Ref.~\cite{Hernandez:2019qed}), whereas it is $a \simeq 0.065\text{ fm}$ for ``B'' ensembles. We provide the pion mass in the valence sector, $aM^\text{v}_\pi$, and the PCAC mass, $am^\text{v}_{\text{pcac}}$. We also include the results for the ratios in Eq.~\eqref{eq:ratios}, and in the last column, the chiral parameter $\xi \equiv {M_\pi^2}/{(4\pi F_\pi)^2}$. Moreover, $\xi_L$ labels $\xi$ corrected by finite-volume effects as explained in the main text. }
\label{tab:ensembles}
\end{table*}

\subsection{Comments on systematics \label{sec:syst}}

We conclude this section by discussing the systematic errors that can affect our results.

We start with finite-volume effects. 
Our ensembles have $M_\pi L > 3.8$ in all cases so we expect finite-volume effects to be small, and suppressed as $1/N_c$. Still, we find that for the observable $\xi$ they can be of $O(1\%)$ and thus we correct for them, as explained in Ref.~\cite{Hernandez:2019qed}, following Refs.~\cite{Gasser:1986vb, Colangelo:2005gd}. 

Since $B_K$ and $\bar{R}^+$ differ by a volume-independent proportionality factor, we can use the results in Ref.~\cite{Becirevic:2003wk}, where the finite-volume effects of $B_K$ have been calculated. In addition, it is known that the finite-volume and chiral corrections of $\bar{R}^+$ and $\bar{R}^-$ are fully anticorrelated \cite{Hernandez:2006kz}. Thus, we find:
\begin{equation}
\bar{R}^\pm(L) = \bar{R}^\pm \left[1 \pm 6 \sqrt{2 \pi } \xi \frac{e^{- M_\pi L}}{(M_\pi L)^{3/2}}(M_\pi L - 4) \right].
\end{equation}
The correction for these quantities is numerically negligible for our ensembles.
While additional finite-volume effects could be present (see Ref.~\cite{Colangelo:2005gd}) we observe that a factor of two increase or decrease of these finite-volume corrections alters  our results  well within the statistical precision.

Concerning discretization effects, we have included the results from two ensembles with a finer lattice spacing at $N_c=3$. Assuming $O(a)$ improvement, we expect that the finer lattice spacing should reduce by $\sim 30\%$ the $O(a^2)$ discretization effects. We observe no significant difference for these data points in Fig.~\ref{fig:fitschiral}, so we see no sign of sizeable discretization errors within our statistical uncertainty. We stress however that a more extensive study is needed for a robust estimate of the discretization error. 

The largest systematic error that we have found is related to the renormalization constants, which we have estimated by one-loop perturbation theory.  We have first compared the non-perturbative renormalization constants of Ref.~\cite{Carrasco:2015pra} to the one-loop perturbation theory results in their setup (they used $c_{sw} =0$). The difference is roughly $\sim 5\%$  for $N_c=3$. On the other hand, we have computed the ratios using $c_{\rm\scriptscriptstyle sw}=1.69$ in the valence sector for the 3A10 ensemble. Using the perturbative renormalization constants for this new value of $c_{\rm\scriptscriptstyle sw}$ we get a result that differs from  our $c_{\rm\scriptscriptstyle sw}=0$ result by roughly $20\%$ in the ratio. Since it is unlikely that this effect can be accounted for by discretization effects, given the tests in a finer lattice mentioned above,  we conclude that there must be significant non-perturbative effects on renormalization constants for the larger
 $c_{\rm\scriptscriptstyle sw}$ (the perturbative one-loop corrections are also significantly larger for the larger value of $c_{\rm\scriptscriptstyle sw}$). This is a large error, and probably a conservative estimate, but it is comparable to the statistical error we achieve, as it will be seen later.

\section{Results}
\label{sec:results}

\subsection{ $N_c$ scaling of $K\to \pi$ amplitudes}

The physical amplitudes $A^\pm$ can be obtained, as explained in Eq.~(\ref{eq:apm}), from  the bare ratios in Table~\ref{tab:ensembles}, and the renormalization coefficients in Tables \ref{tab:renorm} and \ref{tab:renormquenched}. 
As explained 
above, a rigorous way to isolate the (anti-)correlated contributions to the ratios consists on taking the half-sum and half-difference of the ratios. By doing so, the two contributions can be fitted independently since:
\begin{align}
\begin{split}
\frac{{A}^- + {A}^+}{2} &= 1 +\tilde{c} \, \frac{1}{N_c^2}  + \tilde{d} \, \frac{N_f}{N_c^3}  + \hdots ,\\
\frac{{A}^- - {A}^+}{2} &= -\tilde{a} \, \frac{1}{N_c}- \tilde{b} \, \frac{N_f}{N_c^2}  + \hdots. \label{eq:semiR}
\end{split}
\end{align} 

\begin{figure*}[tpb]
   \centering
   \subfigure[ ]%
             {\includegraphics[width=0.475\textwidth,clip]{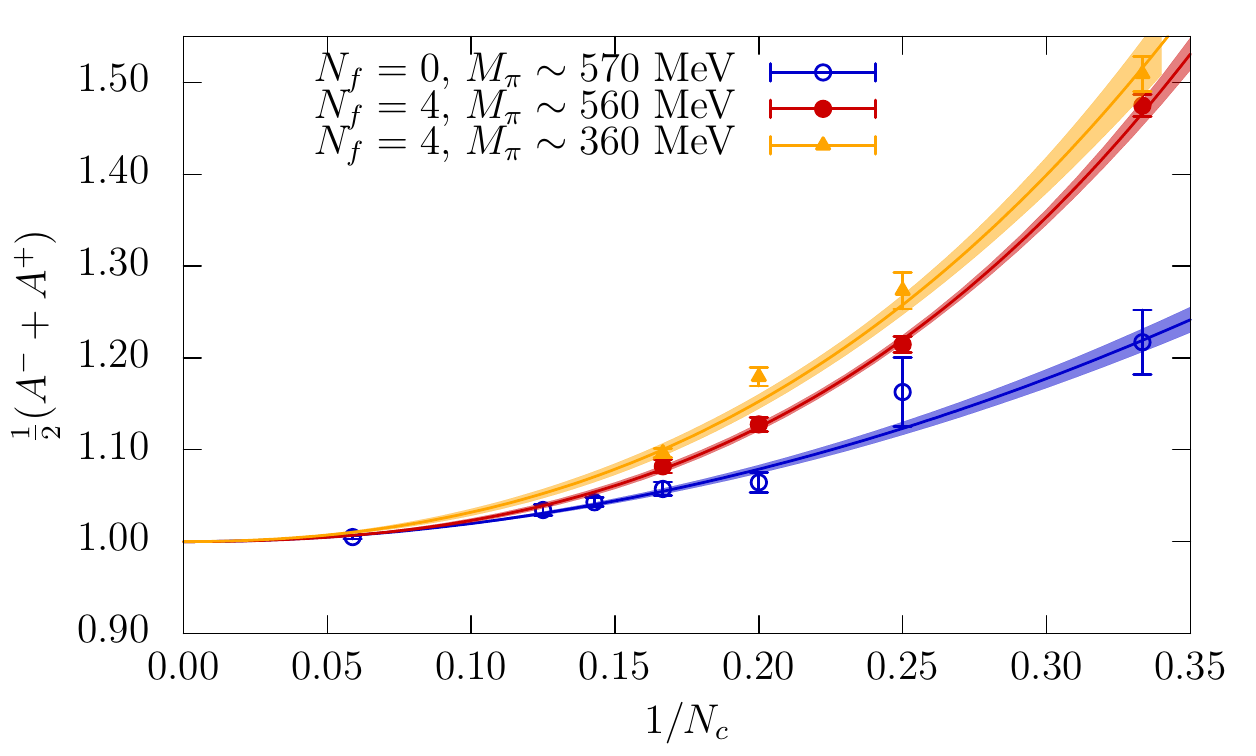}}\hfill
   \subfigure[ ]%
             {\includegraphics[width=0.475\textwidth,clip]{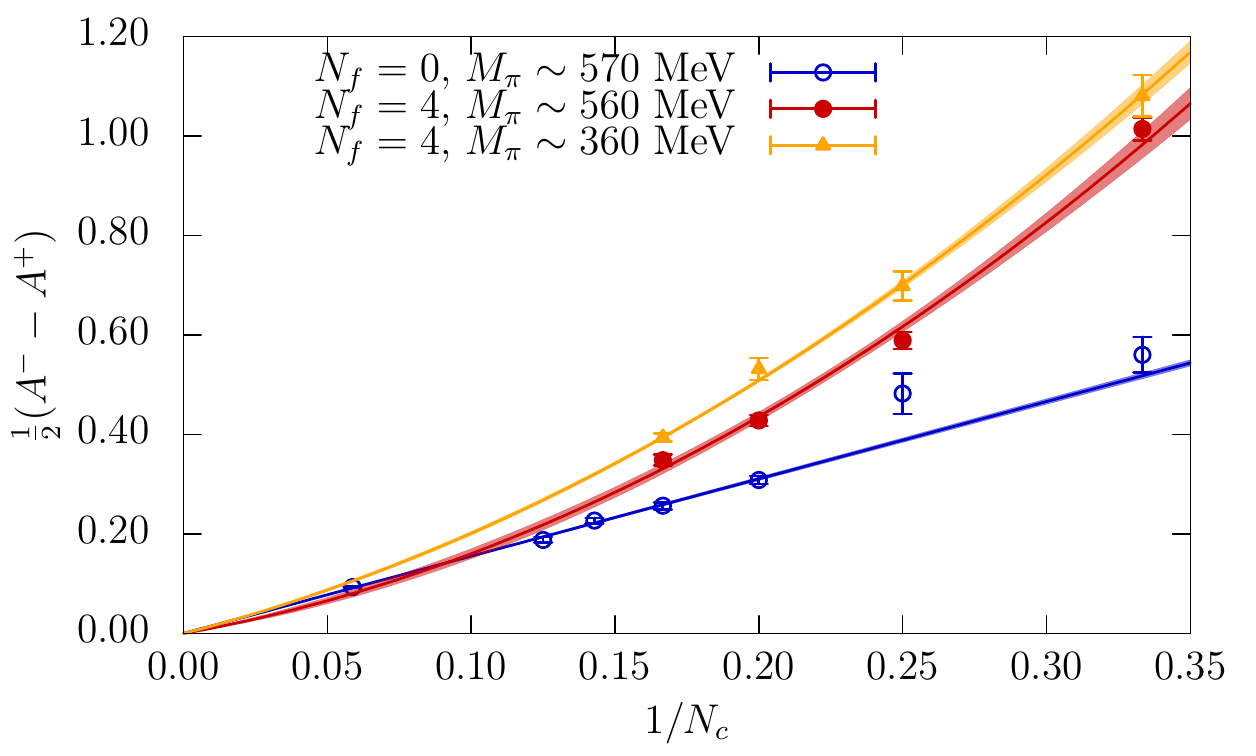}}
   \caption{Half-sum and half-difference of the amplitudes $A^\pm$ as a function of $N_c^{-1}$ for three different cases: (i) quenched results from Ref.~\cite{Donini:2016lwz} in blue, (ii) new dynamical results at a pion similar to the quenched case (red), and (iii) dynamical results at a lighter pion mass (orange). The fit results are shown in Table~\ref{tab:fitsNc}.  Error bars include only statistical errors.}
\label{fig:fitsNc} 
\end{figure*}

In the following, we compare the results of the fits to Eq.~(\ref{eq:semiR}) in three different scenarios:
\begin{enumerate}
\item Quenched results ($N_f=0$) at a heavy pion mass $\sim 570 $ MeV.
\item Dynamical results ($N_f=4$) at a heavy pion mass $\sim 560 $ MeV (ensembles A10).
\item Dynamical results ($N_f=4$) at a lighter pion mass $\sim 360 $ MeV (ensembles A40).
\end{enumerate}
The results for the coefficients $\tilde{a}-\tilde{d}$ for the three scenarios are presented in Table~\ref{tab:fitsNc} and Fig.~\ref{fig:fitsNc}. The coefficients are all of ${\mathcal O}(1)$ and therefore of natural size. Importantly the sign of the $\tilde{a}$ and $\tilde{b}$ coefficients is the same and negative. This implies both terms contribute to reduce the $A^+$ amplitude and enlarge, in a correlated way, the amplitude $A^-$. The fact that $\tilde{b}, \tilde{d} \sim {\mathcal O}(1)$ implies 
 a very large unquenching effect in the large-$N_c$ scaling, and the ratio $A^-/A^+$, which is however compatible with the expansion in 
Eq.~(\ref{eq:semiR}). Specifically, it is due to $\tilde{b}$ and $\tilde{d}$ being absent for $N_f=0$. 
The other two coefficients, $\tilde{a}$ and $\tilde{c}$, are comparable in size in the quenched and dynamical theories. We note however that uncertainties only include statistical errors, and relative discretization errors and the systematics of the perturbative renormalization constants may be significant.
Finally, we observe that  the mass dependence for the $N_f=4$ results seems to affect mostly the coefficient $\tilde{a}$, which is consistent with the chiral dependence in 
Eq.~(\ref{eq:ratioschiral}), and goes also in the direction of enhancing the ratio $A^-/A^+$ towards the chiral limit. 

\begin{table}[h!]
\centering
\begin{tabular}{c|c|c|c|c}
\multicolumn{5}{c}{ Half-difference} \\ \hline
    \rule{0pt}{10.5pt}    Case            & $M_\pi$      & $\tilde{a}$ & $\tilde{b}$ & $\chi^2/\text{d.o.f.}$ \\ \hline\hline
$N_f=0$& $570$ MeV  &  -1.55(2)           &      ---       &8.8/6   \\ \hline
$N_f=4$& $560$  MeV &  -1.03(13)           &   -1.44(13)       & 6.6/2     \\ \hline
$N_f=4$& $360$ MeV &     -1.49(15)        &   -1.32(18)      & 0.3/2      \\ \hline
\end{tabular}

\vspace{0.5cm}

\begin{tabular}{c|c|c|c|c}
\multicolumn{5}{c}{ Half-sum} \\  \hline
  \rule{0pt}{10.5pt}    Case           & $M_\pi$      &$\tilde{c}$ & $\tilde{d}$& $\chi^2/\text{d.o.f.}$ \\ \hline\hline 
$N_f=0$& $570$   MeV&     2.1(1)       &     ---      & 3.5/6  \\ \hline
$N_f=4$& $560$   MeV&\  \ 1.2(3) \  \       &\  \ 2.2(3) \ \ &    1.3/2     \\ \hline
$N_f=4$& $360$  MeV& \  \  2.4(4) \   \    &\  \  1.6(4) \ \ &  3.2/2      \\ \hline
\end{tabular}

\caption{Summary of results for the $1/N_c$ fits to the half-sum and half-difference of the amplitudes $A^\pm$. Errors are only statistical.}
\label{tab:fitsNc}
\end{table}

\subsection{Kaon B-parameter ($B_K$)} 
\begin{figure}[h!]
  \centering
  \includegraphics [width =1 \linewidth]{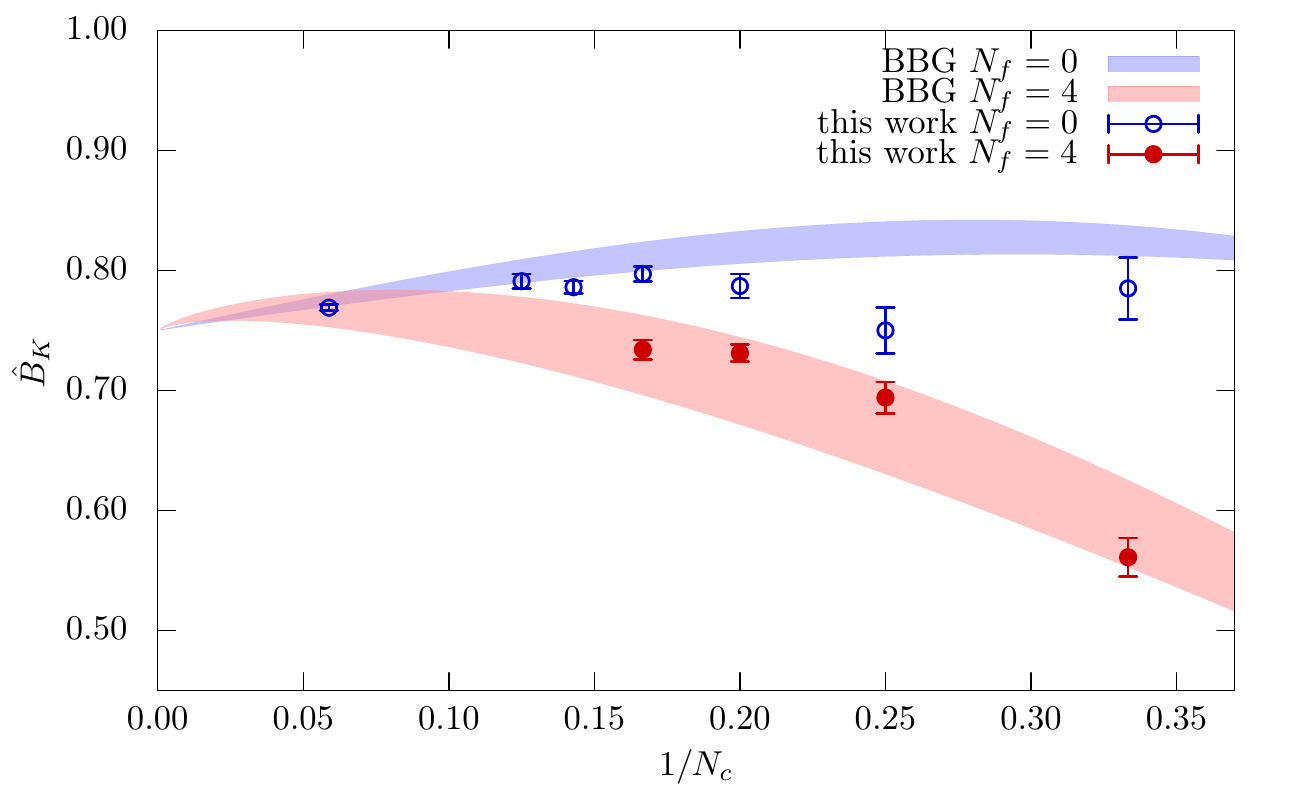}
  \caption{ Lattice results for $\hat{B}_K$, defined in Eq.~(\ref{eq:Bk}), in the case of $N_f=0$ (see Refs.~\cite{Donini:2016lwz, Donini:2017rzi}), and $N_f=4$ (this work). Error bars are only statistical errors. We also include the predictions from Ref.~\cite{Buras:2014maa}, where the band indicates the values obtained when varying the involved matching scale $M$ from 600 to 1000 MeV. \label{fig:Bk} }
 \end{figure}  
     
\begin{figure*}[tpbh!]
   \centering
   \subfigure[ ]
                {\includegraphics[width=0.475\textwidth,clip]{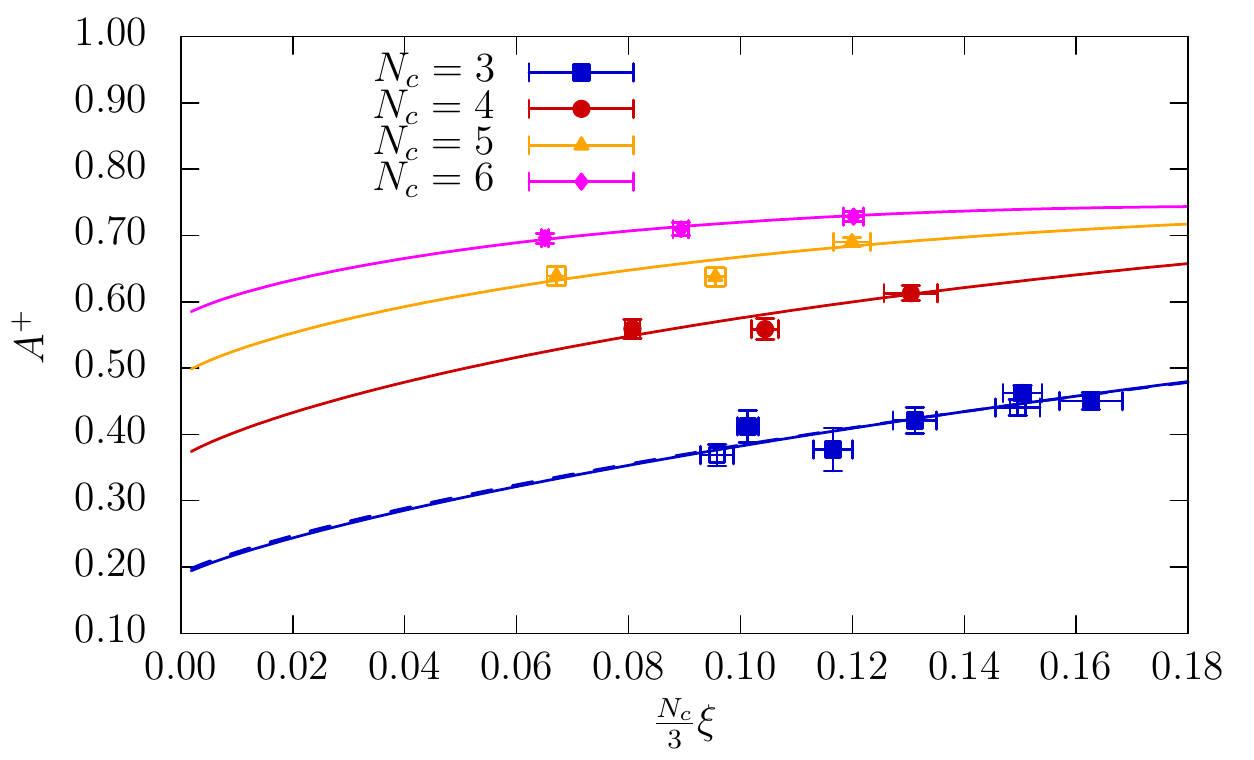}}\hfill
  \subfigure[ ]%
            {\includegraphics[width=0.475\textwidth,clip]{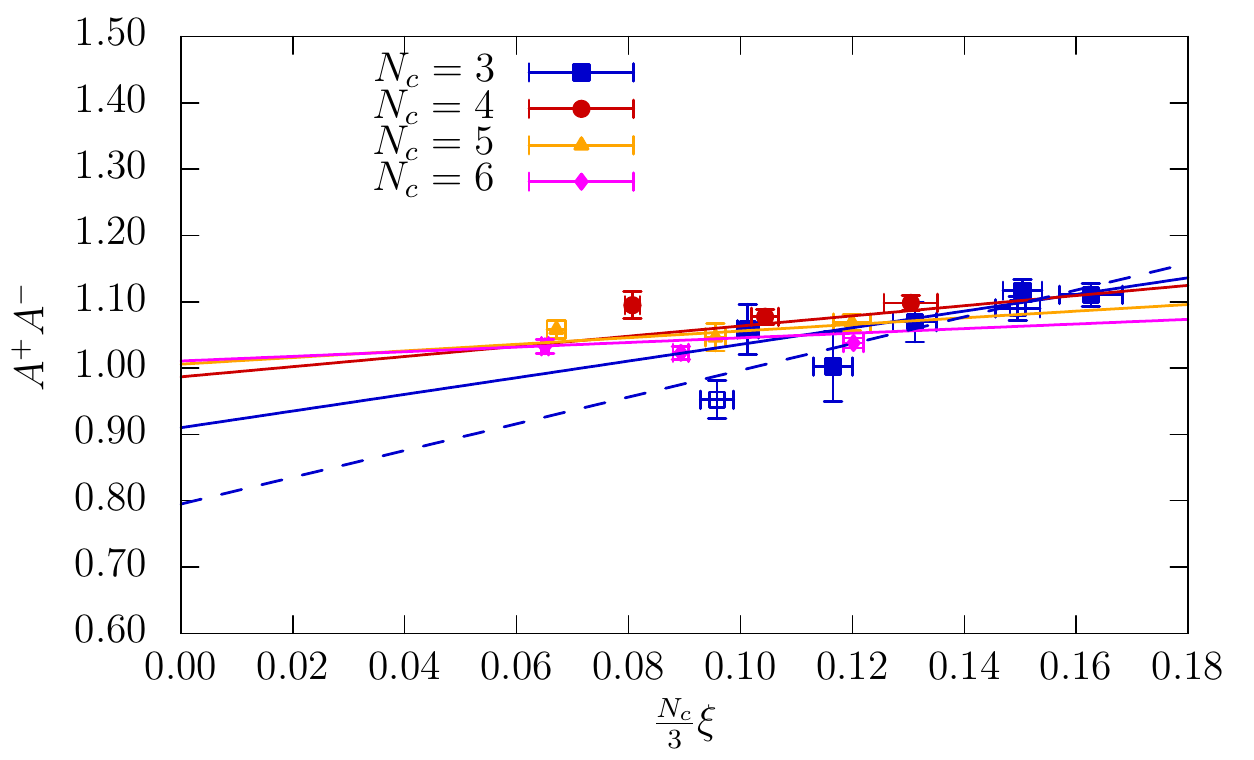}}
   \caption{ Chiral extrapolation of $A^+$ and the product $A^+ A^-$. The data points are also shown in Table~\ref{tab:ensembles}. Empty squares for $N_c=3$ indicate a finer lattice spacing. Solid lines indicate a simultaneous chiral and $N_c$ fit as in Eq.~(\ref{eq:ratioschiral}). Dashed lines represent the chiral extrapolation of the data points for $N_c=3$ following Eqs.~(\ref{eq:ratioschiral}) and (\ref{eq:fitprod}). Errors are only statistical.  \label{fig:fitschiral}  }
\end{figure*}      

The kaon $B$-parameter, $B_K$, is defined from the matrix element of the $\Delta S=2$
operator that mediates neutral kaon oscillations at physical kinematics:
\begin{equation}
\braket{\bar{K}^0| O^{\Delta S=2}(\mu)   | K^0} = \frac{8}{3} f_K^2 M_K^2  \bar B_K(\mu).
\end{equation}
It is customary to quote the renormalization group independent (RGI) version, labelled as $\hat{B}_K$. Its value at the physical point has been computed accurately in $N_f=2$, $2+1$, and $2+1+1$ simulations \cite{Carrasco:2015pra,Durr:2011ap,Laiho:2011np,Blum:2014tka,Jang:2015sla,Bertone:2012cu} (see Ref.~\cite{Aoki:2019cca} for a review).

In our  setup, $\hat{B}_K$ coincides with the renormalized ratio $\bar{R}^+$ up to a normalization. Specifically, we have
\begin{equation}
\hat{B}_K = \frac{3}{4} \hat{c}^+(a^{-1}) \bar{R}^+ \label{eq:Bk}
\end{equation}
where $\hat{c}^+$ can be read off Table~\ref{tab:renorm}.
There are two essential differences in our setup: all meson masses are degenerate, in particular $M_K = M_\pi$, and we have an active light charm quark. Both can significantly affect the value of $\hat{B}_K$. 

We show our results in Fig.~\ref{fig:Bk}. We observe a very significant $N_c$ dependence of $\hat{B}_K$ for $N_f=4$, and a much milder one for $N_f=0$. For $N_c=3$, the quenched result agrees with the standard value of $\hat{B}_K$, while the $N_f=4$ result is about 25\% smaller. We have included as bands the Buras-Bardeen-Gerard (BBG) Dual QCD prediction from Ref.~\cite{Buras:2014maa}, using inputs on meson masses from our own simulations in both cases --- quenched and dynamical. We find that our results are reasonably compatible with the BBG prediction, in particular regarding the suppression of $\hat{B}_K$ in the presence of a light charm.  

To conclude this subsection, we can use the scaling in $N_c$ to infer a value of $\hat B_K$ with three active flavours and quasi-physical kinematics. For this, we use the coefficients $\tilde{a}-\tilde{d}$ in Table~\ref{tab:fitsNc} for the case of $N_f=4$ and $M_\pi=560$ MeV, and so predict the value of $A^+$ with $N_c=3$ and $N_f=3$ at the same value of the pion mass, degenerate with the kaon. We can the get the RGI value $\hat{B}_K$ as in Eq.~(\ref{eq:Bk}), extracting $\bar R_+$ and using the ${\hat c}^+(a^{-1})$ for three-flavour QCD \footnote{ The required parameters  for $N_c=3$, $N_f=3$ are $k^+(M_W) =1.038$, $U^+(a^{-1},M_W)=0.851$, and  $\hat c^+(a^{-1}) =0.841$. In the evaluation of  $\hat{c}^\sigma(a^{-1})$ we have used $\Lambda_{\overline{\text{MS}}}=341 $ MeV from Ref. \cite{Bruno:2017gxd}.  }.  We find
\begin{equation}
\hat{B}_K\big |_{M_K = M_\pi} = 0.67(2)_{\text{stat}} (6)_{Z^+}  (3)_{\text{fit}}  \, ,
\end{equation}
including statistical error, and a $\sim 10\%$ error due to the systematics of the renormalization constants. We also quote a ``fit'' error that we estimate by using the $N_c$ scaling derived from a direct fit of the half-sum and difference of $\bar R^\pm$ instead of $A^\pm$.

We have not found results in the literature for the degenerate case that we can compare to. On the other hand, ChPT relates the value of $\hat{B}_K$ in the degenerate case, to the quasi-physical (QP) situation with $M_\pi=0$ and $M_K$ at its physical value:
\begin{equation}
\hat{B}^{QP}_K = \hat{B}_K\big |_{M_K = M_\pi} \left[ 1 + \frac{2}{3} \left( \frac{M_K}{4 \pi F_K}\right)^2 \log \frac{\Lambda_{\rm eff}^{B_K}}{M_K} \right],
\end{equation}
where $\Lambda_{\rm eff}^{B_K}$ labels an unknown scale that parametrizes the effect of the unknown LECs. For $\Lambda^{B_K}_{\rm eff}>M_K$,   $\hat B^{QP}_K$ is larger than 
$\hat{B}_K$ and could  be compatible with the existing results at the physical point from $N_f=2+1,N_c=3$ simulations \cite{Carrasco:2015pra,Durr:2011ap,Laiho:2011np,Blum:2014tka,Jang:2015sla,Bertone:2012cu}.

\subsection{ Extraction of the effective couplings $g^\pm$ }

The main goal of this work is to compute the ratio $g^-/g^+$ by extrapolating $A^\pm$ to the chiral limit. For the required chiral extrapolation, we follow the same strategy as in 
Ref.~\cite{Giusti:2006mh}. We extract $g^+$ from a chiral fit to $A^+$, and the product $g^+ g^- $ from that of the product $A^+ A^-$ . The ratio can then be evaluated as
\begin{equation}
\frac{g^-}{g^+} \equiv \left(g^- g^+ \right) \times \frac{1}{(g^+)^2} \, .
\end{equation}
This approach results in a milder chiral extrapolation, that will hopefully introduce a smaller systematic error.

We have performed two kinds of fits. In Fit 1, we use all data points with $N_c=3-6$ in a simultaneous chiral and $N_c$ fit using Eqs.~(\ref{eq:ratioschiral}) and (\ref{eq:fitprod}), incorporating the $1/N_c$ expansion of the couplings as in Eqs.~(\ref{eq:gNc},\ref{eq:LNc},\ref{eq:Ncprod}). In Fit 2, we fit using only the data with $N_c=3$, and extract the effective couplings for this theory. This way, for $N_c=3$ we find:
\begin{align}
\begin{split}
\text{Fit 1: } &  g^+ = 0.187(21),\ \ \ g^+ g^- = 0.91(4), \\
\text{Fit 2: } &  g^+ = 0.190(27),\ \ \ g^+ g^- = 0.80(6).
\end{split}
\end{align}
The complete results of these fits are shown in Tables~\ref{tab:fitschiralNc}, and \ref{tab:fitsNc3}, and also in Fig.~\ref{fig:fitschiral}. 

From these results, we obtain for the ratio of couplings at $N_c=3$:
\begin{equation}
\frac{g^-}{g^+} \Bigg |_{\text{fit 1}} = 26(6), \ \ \ \ \frac{g^-}{g^+} \Bigg |_{\text{fit 2}} = 22(5), \label{eq:fitres}
\end{equation}
where errors are only statistical, but correlations are taken into account.

\begin{table}[h!]
\centering
\begin{tabular}{c | c | c | c | c | c}
\multicolumn{6}{c}{Fit  1 for $A^+$} \\ \hline 
  \rule{0pt}{10.5pt}   $a_\chi$             &  $N_f b_\chi + c_\chi$          & $N_f{d}_\chi$ & $L^{(0)}$ & $L_+^{(1)}$  &  $\chi^2/\text{d.o.f.}$  \\ \hline 
-2.2(6) & -3(4)  & 7(7) &2.4(8) &-11(4) &    12.0/11  \\ 
 \hline  
\end{tabular}

\vspace{0.5cm}
\begin{tabular}{c | c | c | c     } 
\multicolumn{4}{c}{Fit 1  for $A^+ A^-$} \\ \hline 
 \rule{0pt}{10.5pt}  $\alpha$          & $\beta$ & $L_-^{(1)} - L_+^{(1)}$   &  $\chi^2/\text{d.o.f.}$ \\ \hline 
 1.6(4) &-7.2(9) & 1.4(4)  & 26.7/13   \\ 
 \hline
\end{tabular}
\caption{Results for Fit 1:  the simultaneous chiral and $N_c$ fits  for $A^+$ and $A^+ A^-$. Errors are only statistical.}
\label{tab:fitschiralNc}
\end{table}

\begin{table}[h!]
\centering
\begin{tabular}{c |c | c }
\multicolumn{3}{c}{Fit 2 for  $A^+$} \\ \hline
  \rule{0pt}{10.5pt}   $g^+$         & $L_+^r$  &   $\chi^2/\text{d.o.f.}$ \\ \hline
0.190(27) & -1.1(7) &  4.9/5    \\  \hline 
\end{tabular}

\vspace{0.5cm}
\begin{tabular}{c | c | c   } 
\multicolumn{3}{c}{Fit  2 for $A^+ A^-$} \\ \hline 
 \rule{0pt}{10.5pt}   $g^+ g^-$ & $L_-^r - L_+^r$   &  $\chi^2/\text{d.o.f.}$\\ \hline
0.80(6) & 0.8(2) & 6.2/5  \\ \hline
\end{tabular}
\caption{Results for Fit 2: the chiral fit at $N_c=3$ for $A^+$ and $A^+ A^-$. Errors are only statistical.}
\label{tab:fitsNc3}
\end{table}

\subsection{$K \to \pi\pi$ amplitudes in ChPT}

\begin{figure}[h!]
  \centering
  \includegraphics [width =1 \linewidth]{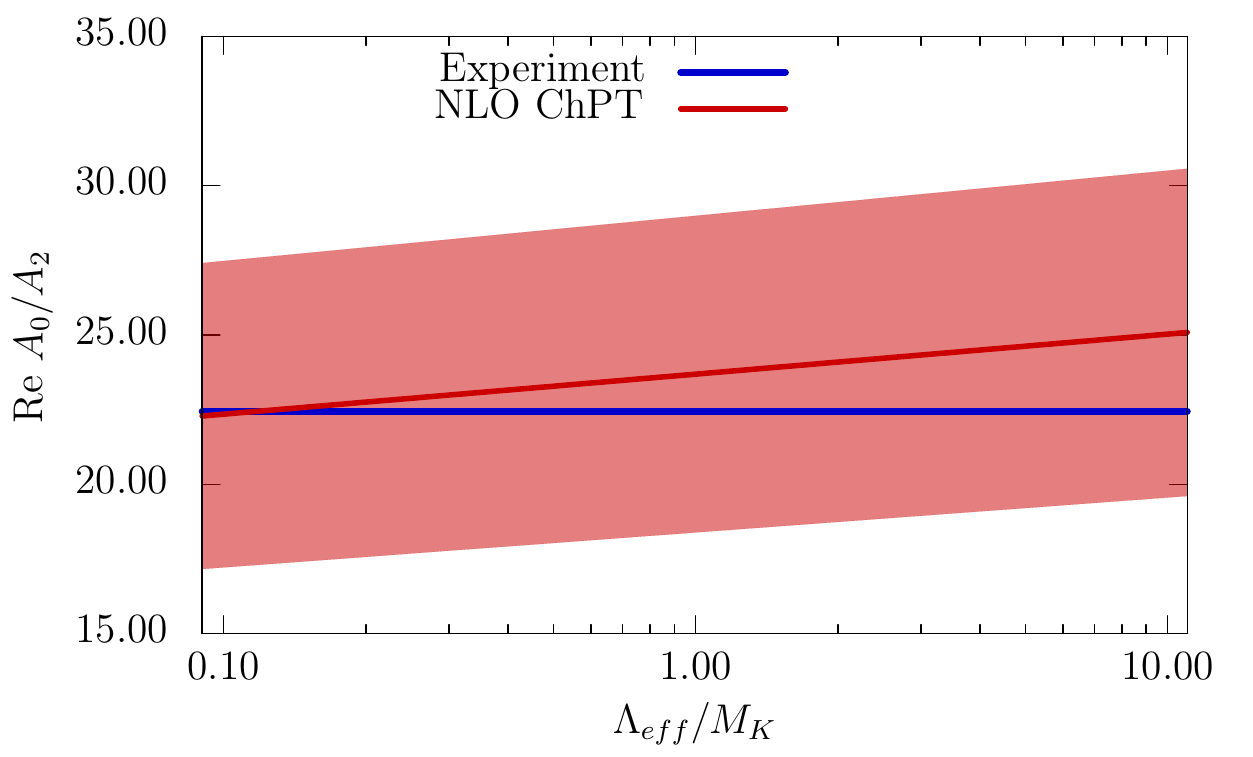}
  \caption{ NLO ChPT prediction (in red) for the ratio of $K \to \pi\pi$ isospin amplitudes as a function of the NLO LEC, $\Lambda_{\rm eff}$. We use the input of Fit 2 in Eq.~\eqref{eq:fitres}. This prediction is valid for $M_\pi=M_D =0$, and $M_K$ at its physical value. The shaded area represents the statistical error associated to the  ratio of couplings --- see Eq.~\eqref{eq:fitres}. As a guideline, we also show the experimental value for the ratio of amplitudes (in blue).  \label{fig:ktopipi} }
 \end{figure}

Using the result for the ratio of couplings in Eq.~\eqref{eq:fitres}, and the NLO ChPT prediction in Eq.~(\ref{eq:NLOktopipi}), we can obtain an indirect result for the ratio of isospin amplitudes in the $K \to \pi\pi$ decay for $N_c=3$. In Fig.~\ref{fig:ktopipi}, we show this prediction as a function of an unknown effective scale $\Lambda_{\rm eff}$. This prediction, valid for $M_\pi=M_D=0$ and physical $M_K$, shows small NLO effects in a wide range of values of the effective scale. 

We are now in the position to quote a final result for the ratio of isospin amplitudes:
\begin{equation}
\text{Re }\frac{A_0}{A_2} \Bigg \rvert_{N_f=4} = 24(5)_{\text{stat}}(4)_{\text{fit}}(5)_{Z^\pm}(3)_{\text{NLO}},
\end{equation}
where the central value comes from the fit 2 result in Eq. \eqref{eq:fitres}. In the previous equation, the various error sources originate as follows : (i) statistical error, (ii) systematic error from the difference between fit 1 and 2 in Eq.~\eqref{eq:fitres}, (iii) a $20 \%$ error from the renormalization constants --- see Section \ref{sec:syst} ---, and (iv) a $10\%$ error from the NLO effects --- see Fig.~\ref{fig:ktopipi}.  Combining all error sources in quadrature results in a $\sim 30 \%$ uncertainty on the total result,
which is dominated by systematics.
We also stress that this is a result in the theory with a light charm quark. Interestingly, this indirect computation yields a value compatible with the experimental result for the $\Delta I=1/2$ enhancement.

\section{Conclusions}
\label{sec:conclusions}

We have presented the first non-perturbative study of the scaling of $\Delta S=1$ weak amplitudes with the number of colours, $N_c=3-6$, in a theory with four degenerate light flavours $N_f=4$.  These results  have been obtained from dynamical simulations with clover Wilson fermions, at $a\simeq 0.075$ fm and $a\simeq 0.065$ fm and pion masses in  the range $360-570$ MeV. We have  analysed the $K\to\pi$ amplitudes $A^\pm$, mediated by the two current-current operators $Q_\pm$ of the $\Delta S=1$ weak Hamiltonian in Eq.~(\ref{eq:heffs1}).

The diagrammatic analysis of the large-$N_c$ scaling of these observables presented in Sect.~\ref{sec:scaling} allows to classify the subleading $N_c$ corrections, and demonstrates the anticorrelation of the leading ${\mathcal O}(1/N_c)$ and ${\mathcal O}(N_f/N^2_c)$ contributions in the $A^\pm$ amplitudes. Our numerical results confirm this expectation and show that these
corrections are naturally large in the Veneziano scaling limit, i.e., the coefficients of both corrections are ${\mathcal O}(1)$. They can nevertheless explain the large enhancement of the ratio $A^-/A^+$ for  $N_c=3$ with respect to the $N_c\rightarrow \infty$ limit. This involves an unprecedentedly large unquenching effect in this ratio, that is nevertheless compatible with natural size ${\mathcal O}(N_f/N_c^2)$ corrections.

The amplitudes $A^\pm$ in the chiral limit can be matched to their ChPT counterparts,  which depend on the leading low-energy couplings, $g^\pm$, of the chiral effective weak Hamiltonian. From a chiral extrapolation of the combinations $A^+$ and $A^+ A^-$, we have then extracted the couplings $g^\pm$, which are finally used to predict in ChPT the ratio of $K \to (\pi\pi)_{I=0,2}$ amplitudes. In particular, we have obtained an indirect prediction of the ratio of isospin amplitudes, $A_0/A_2$, by this procedure which seems to largely account for the elusive ``$\Delta I=1/2$ rule''.  Our estimate for this ratio in the theory with a light charm is
\begin{equation}
\text{Re }\frac{A_0}{A_2}  \Bigg \rvert_{N_f=4}  = 24(5)_{\rm\scriptscriptstyle stat}(7)_{\rm\scriptscriptstyle sys},
\end{equation}
which suggests that the enhancement may indeed be largely dominated by intrinsic QCD effects. 
\\
\begin{acknowledgments}
We thank the HiRep developers for providing us with a $SU(N_c)$ lattice code, particularly C.~Pica and M.~Hansen. We acknowledge useful discussions with M. Garc\'ia P\'erez, A. Gonz\'alez-Arroyo, 
G.~Herdo\'iza, A. Pich, A.Ramos, A. Rago, S. Sharpe, and C. Urbach.

This work was partially supported through the Spanish MINECO-FEDER projects FPA2015-68541-P, FPA2017-85985-P and PGC2018-094857-B-I00 and the Centro de Excelencia Severo Ochoa Programme SEV-2016-0597, as well as the European projects H2020-MSCAITN-2018-813942 (EuroPLEx), H2020-MSCA-ITN-2015/674896-ELUSIVES, H2020- MSCA-RISE-2015/690575-InvisiblesPlus and STRONG-2020 (under grant agreement No 824093). Finally we acknowledge partial support from the Generalitat Valenciana  grant PROMETEO/2019/083.

The work of FRL has also received funding from the European Union Horizon 2020 research and innovation program under the Marie Sk\l{}odowska-Curie grant agreement No. 713673 and ``La Caixa'' Foundation (ID 100010434). 

We acknowledge the computational resources provided by Cal\'endula (SCAYLE), Finis Terrae II (CESGA), Mare Nostrum 4 (BSC), Lluis Vives (UV) and Tirant III (UV).
\end{acknowledgments}

\vspace{0.5cm}

\bibliography{biblio.bib}

\end{document}